# Sky maps without anisotropies in the cosmic microwave background are a better fit to WMAP's uncalibrated time ordered data than the official sky maps

Short title: The WMAP uncalibrated data is a better fit to no anisotropies

Keith S Cover

Corresponding author:
Keith S Cover, PhD
Department of Physics and Medical Technology
VU University Medical Center
Postbus 7057
1007 MB Amsterdam
The Netherlands
Email: Keith@kscover.ca

## Abstract

The purpose of this reanalysis of the WMAP uncalibrated time ordered data (TOD) was two fold. The first was to reassess the reliability of the detection of the anisotropies in the official WMAP sky maps of the cosmic microwave background (CMB). The second was to assess the performance of a proposed criterion in avoiding systematic error in detecting a signal of interest. The criterion was implemented by testing the null hypothesis that the uncalibrated TOD was consistent with no anisotropies when WMAP's hourly calibration parameters were allowed to vary. It was shown independently for all 20 WMAP channels that sky maps with no anisotropies were a better fit to the TOD than those from the official analysis. The recently launched Planck satellite should help sort out this perplexing result.

**Introduction.** - The measurement of anisotropies in the cosmic microwave background (CMB) is of great interest to astronomers. The sky maps generated by the WMAP mission are widely considered to be the best measurements of the reported anisotropies [1-3]. The measurements of the CMB made by the WMAP satellite did not directly measure the intensity of the CMB at each point in the sky. Rather, each sky map of the CMB was reconstructed from hundreds of millions of differential measurements, usually referred to as the time ordered data (TOD). Image reconstruction, including calibration of the TOD, was used to reconstruct the sky maps from the TOD.

The CMB is widely believed to have been radiated by the universe when it was about 400,000 years old. Therefore, any variation in the intensity of the CMB with direction in the sky would provide priceless information on the early universe. The CMB has been shown to have a nearly perfect black body spectrum with a temperature of 2.726K [4]. Black bodies with this temperature radiate in the 60-630GHz range with a peak frequency of about 160GHz. Also, the CMB has a nearly uniform intensity with direction.

*The WMAP sky maps.* – A sky map is an intensity plot of radiation in all directions at a particular frequency (Fig1). At the galactic equator is the signal generated by the dust and gas in our galaxy that obscures the CMB that is called the galactic band. Our solar system's relative motion to the CMB of about 369km/s introduces a small but smooth variation in the intensity of the CMB, usually referred to as the dipole, at about 1 part in 10,000 [1]. The exact intensity of the dipole can be predicted [5] and easily subtracted off the CMB. Any variations in the intensity of the CMB after the subtraction are referred to as anisotropies.

The reported anisotropies have intensity variations roughly 10 to 20 times smaller than those of the dipole. Thus, highly sensitive and well calibrated instrumentation is needed to reliably measure the reported anisotropies. Figure 1 shows the sky map generated by the official WMAP analysis of the 5 year data for the 94GHz frequency band both with and without the dipole subtracted. For display purposes only, the sky map has been averaged down from WMAP resolution to COBE resolution. Each COBE pixel is the average of about 500 WMAP pixels.

While there have been many reports of the detection of anisotropies in the literature, the first widely accepted detection of the anisotropies used the COBE satellite and was reported by Smoot et al. [6]. The WMAP mission [1-3], which was launched 10 years after COBE, reported anisotropies with similar intensity while resolving them with much higher spatial resolution.

Since the initial reported detection by COBE, many experiments have reported the detection of anisotropies in the CMB consistent with the COBE results (see Bennett el al. [1] for an extensive list). The WMAP measurements of the CMB are widely considered to be of far better quality than any previously acquired measurements. In addition, the WMAP TOD, along with some of the software used in processing the TOD, is publicly available and well documented. Therefore, only the uncalibrated TOD from WMAP will be reanalysed in this paper.

The central concern of this paper is the initial calibration of WMAP's TOD. For perfectly calibrated TOD, the WMAP reconstruction problem becomes linear, and as the literature shows, there is a robust and reliable reconstruction algorithm [7-9]. However, the 20 independent channels on WMAP each have drifts in both gain and baseline that approach 5mK over a year [2]. This drift is more than 10 times the level of the reported anisotropies. Thus, if the linear reconstruction algorithm was applied to the uncalibrated TOD, the drift of the channels would swamp the anisotropies preventing their detection. To date, the only publications to consider the calibration of WMAP's TOD are those by the WMAP team [2, 3]. Since the proper calibration of the WMAP TOD's is critical to reliable detection of the anisotropies, it deserves more extensive study.

The WMAP image reconstruction algorithms were designed and implemented by very competent and well funded researchers [1-3]. Thus any major problems with the image reconstruction of the sky maps are likely to primarily reflect problems with the current state of the art in the design of image reconstruction algorithms. In particular, the calibration of WMAP's uncalibrated TOD used an image reconstruction algorithm based on the maximum likelihood criterion. The maximum likelihood criterion is widely used in the design of image reconstruction algorithms but has a mixed record in terms of the reliability of the algorithms designed.

*The proposed criterion* - By reliable reconstruction it is meant that any signal that stands well above the noise, such as the anisotropies do in the official WMAP sky maps, actually exists, and is not a systematic error or some other artefact of the image reconstruction. The goal of the proposed criterion [10-13] is to provide a statistical test for measured data to ensure that a signal of interest actually exists in the measured data.

The proposed criterion avoids the conundrum of some images consistent with the data having the signal of interest and others not. It accomplishes this by requiring that, for a detection of a signal of interest to be considered reliable, it must be shown that all images consistent with the data must have the signal of interest.

Adherence to the proposed criterion can be tested with a null hypothesis. While the proposed criterion may seem onerous, the highly trusted windowed discrete Fourier transform reconstruction algorithm used in MRI imaging satisfies the criterion and inspired the criterion's design. The proposed criterion has also been applied to multiexponential reconstruction [10, 11, 13] and detection of the anisotropies in the COBE TOD assuming the calibration of the COBE TOD was correct [12]. To fully understand the proposed criterion, it is important to keep in mind the difference between probabilities and probability densities.

In this reanalysis, the reliability of the anisotropies in the WMAP sky maps will be assessed by applying the proposed criterion. Adherence to the proposed criterion will be tested by determining whether WMAP's uncalibrated TOD is a better fit to the reported anisotropies or to no anisotropies.

**Methods.** -The WMAP team has provided detailed descriptions of WMAP's design, performance specifications, data processing pipeline and results in many publications [1-3]. As a brief summary, the WMAP satellite was launched in 2001 with the primary purpose of measuring anisotropies in the CMB. WMAP makes differential measurements of the CMB with its antenna beams separated by $\cong 140°$ as it spins about its axis every 129s. The two beams are referred to as the A and B beams. It simultaneously measures in 5 frequency bands (K, Ka, Q, V, W) with centre frequencies of 23GHz, 33GHz, 41GHz, 61GHz and 94GHz. In total, WMAP has 20 independent channels with more channels at the higher frequencies. WMAP requires one orbit around the sun, taking one year, to acquire sufficient TOD to reconstruct a map of the whole sky. To date, the uncalibrated TOD for the first 5 years of WMAP's data acquisition has been released [3].

Throughout this reanalysis, as in the calibration stage of the official analysis, only a subset of the samples in the uncalibrated TOD was used. Any sample in which either of the A or B beams was pointed at a moving object, such as the moon or Jupiter, was deleted. Also, if either beam A or B was pointed at the galactic band or other known source of microwaves that were not from the CMB, the sample was deleted. This step, referred to in the official analysis as masking, used the mask map_temperature_analysis_mask_r9_5yr_v3.fits, which is available on the official WMAP LAMBA web site. It masked out about 18% of the pixels in each sky map. After masking, the uncalibrated TOD for each channel had about 65% of the samples remaining. The reanalysis was also rerun with another 5 year WMAP mask that masked 28% of the pixels in a sky map to see if more masking had any significant effect on the results. Only the first year of each of the 5 year TOD's were used for this reanalysis.

The Doppler shift signal due to WMAP's motion relative to the CMB is usually broken down into two parts. The first part is the constant signal due to the solar system's barycenter motion relative to the CMB. This signal is usually referred to as the dipole. The second part is the time varying signal due to WMAP's orbit about the sun's barycenter, referred to as the orbital Doppler shift. The value for the dipole used throughout this paper were those for the COBE dipole (-0.2234, -2.2228, 2.5000mK) [1]. The calculations were also repeated using the 5 year WMAP dipole [3] to see if this change yielded any significant differences in the results.

In WMAP's differential design each of the 20 channels consisted of 2 sub channels, referred to as sub channels 3 and 4 [2]. The difference between these two sub channels, after calibration, yields the difference between the intensity of the A and B beams. Applying the correction for the orbital Doppler shift to this difference yields the TOD used in the linear stage of the image reconstruction. The calibration step converts the signal in milliKelvin (mK) to the integer digital units (DU) of the analogue to digital converters. The official analysis used the following equation to calibrate the sub channels for each channel

$$S^{uncal}_{j,n} = g_k S^{cal}_{j,n} + b_k \quad (1)$$

where the index *n* corresponds to the time point of each measurement and j corresponds to sub channel 3 or 4. Here $g_k$ and $b_k$ are the hourly gains and baselines of the sub channels and the index *k* represents the hour of the calibration parameters. This reanalysis also uses equation (1) but varied the hourly calibration parameters for each of the 20 channels to find the best fit to the uncalibrated TOD assuming the null hypothesis that there were no anisotropies.

As with the official analysis, throughout this reanalysis, the fit for each channel of the predicted uncalibrated TOD to the measured uncalibrated TOD was quantified by finding the sum of the squares of the residuals. The residuals were defined to be the difference between the measured and predicted uncalibrated TOD in DU's. As the WMAP TOD is differential, there are two residuals for each time point.

For both the official analysis and this reanalysis, the values of the gain and baseline for each hourly interval (Eq 1) were calculated using simple linear regression [14]. For the linear regression, the dependent variable was the measured uncalibrated TOD in DU and the independent variable was the predicted TOD in milliKelvin. For the null hypothesis, the predicted calibrated TOD consisted only of the orbital Doppler shift plus the dipole. Once the calibration parameters had been calculated, the calibration parameters were then used to convert the predicted TOD from milliKelvin to DU's.

To calculate the predicted calibrated TOD corresponding to the official analysis, first the calibration parameters provided by the WMAP team were used to calculate the sky map, including the reported anisotropies, for each of the 20 channels [7]. The reconstructed sky map was then used to generate an interim predicted TOD in milliKelvin. The orbital Doppler shift was then added to the interim predicted TOD to yield the calibrated predicted TOD. Finally, the predicted TOD in milliKelvin was converted to DU's using the official calibration parameters.

Some simple calculations yield a good estimate of the statistical significance of the difference between the sum of the squares for the official calibration and the calibration using the null hypothesis. Both sums of squares have the same noise since they use the same uncalibrated TOD. For large number of samples, and assuming the noise in the measured uncalibrated TOD is ideal (Gaussian, uncorrelated, stationary, mean of zero, additive), the statistical distribution for $\chi^2$, to a good approximation, is Gaussian. The expected mean and standard deviation of the distribution are then N and sqrt(2N) respectively where N is the number of residuals [14 p 807, 15]. Because of WMAP's differential measurements, N is twice the number of samples in each channels masked TOD. To obtain an measure of the significance of the difference between the sum of squared residuals in the official reconstruction and the sum of residuals under the null hypothesis, we divide this difference by the estimated standard deviation (i.e. by the square root of 2*2N). This significance value provides a statistical significance test for the results for each channel independently.

**Results.** - Table 1 shows a small selection of the calibration parameters provided by the WMAP team and those calculated assuming the null hypothesis. The values for the two are similar. The similarity between the two sets of calibration parameters is largely because the dipole is the dominant signal in both the official and null hypothesis sky maps. Note that the gain for sub channel V113 of the official calibration jumps 4% in as little as 2 hours. The amplitudes of the reported anisotropies are about 5% to 10% of the dipole.

As can be seen in Hinshaw et al. [2], the large fluctuation of the gain from hour to hour was a characteristic of the official calibration. The standard deviation of the noise in the calibrated data fluctuates from hour to hour by up to 5 percent or more as opposed to less than one percent of the whole year for the uncalibrated data.

Table 2 shows the sum of the squares of the residuals for both the official calibration and the null hypothesis calibration. Results are shown for the COBE dipole and the mask that removes 18% of the sky map pixels. For all 20 channels, the measure of significance shows that the fit to the uncalibrated data is significantly better for the null hypothesis than for the official calibration.

Values within two standard deviations of each other are widely considered a statistical tie. However, the smallest number of standard deviations in Table 2 is 127. Thus, for every WMAP channel, the null hypothesis combined with its calibration is a much better fit to the measured uncalibrated TOD than the official sky map combined with its calibration.

The repeat of these calculations for all 20 channels with the dipole from the official WMAP 5 year analysis or with the larger mask yielded little change in significance. For example, for channel V11 the significance was 476 standard deviations compared to 585 for the 5 year dipole. Similarly, using the mask that eliminated 28% of the sky map pixels for channel V11 reduced the significance to 353.

**Discussion. –** The results of this reanalysis of the WMAP uncalibrated TOD show that, when the calibration parameters are allowed to vary, the uncalibrated TOD is a better fit to no anisotropies than the anisotropies reported by the official analysis. As this is the first evaluation of the calibration used by WMAP, other than that provided by the WMAP team, this raises important questions about the reliability of the anisotropies reported by the official WMAP analysis.

The WMAP image reconstruction algorithms were designed and implemented by very competent and well funded researchers [1-3]. Thus any major problems with the image reconstruction of the sky maps are likely to primarily reflect problems with the current state of the art in the design of image reconstruction algorithms. In particular, the calibration of WMAP's uncalibrated TOD used an image reconstruction algorithm base on the maximum likelihood criterion. The maximum likelihood criterion is widely used in the design of image reconstruction algorithms but has a mixed record in terms of the reliability of the algorithms designed.

There are two key characteristics of the official WMAP sky maps that have inspired confidence in the existence of the reported anisotropies. The first characteristic was that the reported anisotropies show the same pattern in the sky maps at other frequency bands. The second characteristic was that the reported anisotropies have the same pattern from year to year.

Therefore, for the reported anisotropies to be, either all or in part, an artefact of the image reconstruction would require the artefact to be both reproducible over frequency bands and from year to year. The Doppler shift due to WMAP's motion relative to the CMB was the same at all frequency bands and from year to year. Since the best estimate of signal due to WMAP's motion relative to CMB was subtracted off as part of the official analysis, any errors in calibration parameters could yield consistent artefacts if done in a consistent way year to year and over frequency bands. This suggests that it is possible that the information that generated the reported anisotropies could have been inadvertently imprinted on the calibration parameters during the calibration process. This is not a possibility that was considered in the WMAP papers. For example, Hinshaw et al. 2003 [2] pg 64 states "Slow drifts are removed as part of the calibration procedure, but signals near the spin period can couple to the sky maps with some efficiency." However, the results of this reanalysis seem to show the calibration parameters can cancel out, or by the same argument, mimic the reported anisotropies, even though they only change hourly.

The WMAP team used slightly different methods to calculate their calibration coefficients in the first, third and fifth year analysis [2, 16, 3]. But common to all of the analyses was an iterative process that used the WMAP's motion relative to the CMB background as the basis of the calibration. The sky maps for each channel were initialised with the COBE dipole. However, the direction and intensity of the dipole was allowed to vary during the iterations. The final value of the dipole for the 5 year analysis, which is the WMAP 5 year dipole mentioned above, are very close to the those of the COBE dipole.

Another common characteristic of the WMAP's team calculation of the calibration coefficients was that all spherical harmonics, other than the dipole, were filtered out during each iteration. The remaining dipole was used to calculate the predicted calibrated TOD for the hourly linear regression. An estimate of the higher order spherical harmonics was also subtracted from the measured uncalibrated TOD before the hourly linear regression. This estimate was generated using the value of the calibration coefficients available during the iteration. The WMAP's team found the calibration coefficients to converge in anefficient and stable manner during the iterations.

A fundamental problem with any estimation of the WMAP calibration parameters is the optimisation problem is a nonlinear one. The possibility of iterative methods converging to local minima, rather than global ones, exists unless explicitly ruled out. Therefore, given the results of this reanalysis it may be worth re-examining the possibility that the official WMAP calibration may have converged on local minima. Given the consistency of the anisotropies over all 20 channels, the channels would have had to have converged to similar minima. This case is within the realm of possibility because all twenty channels have a strong signal from the dipole. And, as was mentioned above, the dipole has many of the characteristics that are considered reliable indicators of the anisotropies.

To simplify and speed up this reanalysis, several steps included in the official analysis were left out. These steps were considered in the official analysis to have only minor effects on anisotropies in the sky maps [2]. For example, no correction was made in this reanalysis for the realistic antenna beam, thermal drifts of the amplifiers or the loss imbalance between sides A and B.

Other concerns about the WMAP sky maps have been raised in the literature because some of their characteristics were judged improbable [17, 18] or as having unexpected correlations [19]. In particular, both the quadrupole and some of the octopoles were found orthogonal to WMAP's orbit around the sun. This alignment was found to have a probability of less than 1 in 100 [20, 21]. When all these factors are combined, there is a case for re-evaluating the calibration of the WMAP TOD while keeping in mind the possibility that the reported anisotropies may be partially or wholly an artefact of WMAP's image reconstruction, including the calibration.

The results of this reanalysis suggest that the official WMAP sky maps are somehow inaccurate. However, this better fit gives little idea as to the characteristics of the inaccuracies. There have been suggestions that, while calibration drift may affect lower order spherical harmonics, it will have far less impact on the higher order ones [3]. This possibility cannot be ruled out based on this reanalysis. However, there has been no rigorous evaluation of this possibility published in the literature.

The Planck satellite will provide the first measurements of the CMB that allow direct detection of the anisotropies because of its use of repetitive circular scans. One minute is the time it takes the Planck satellite to measure a full circle of a repetitive circular scan. So where WMAP has to correct for calibration drift over 1 year, Planck only has to deal with drifts over 1 minute. If the same cycle is repeated over and over for hours or days, the average of the repetitions has a calibration that can be represented by two parameters, its gain and baseline. Thus any anisotropies can be easily detected by plotting the signal over the averaged repetitive circular scan. However, Planck's reliability in detecting anisotropies does not extend to the estimation of the spherical harmonics. As the whole sky must be scanned for Planck to generate reliable estimates of the spherical harmonics, the Planck's calibration drift over a full year becomes an issue, just as with WMAP.

Since the release of the 1st year WMAP results, there have been other ground and balloon based missions that have yielded very similar results to WMAP including Boomerang03 [22]. These missions used more modern and sensitive instrumentation. Calibration of ground and balloon based missions is challenging for many reasons including distortions due to the earth's atmosphere. It would be worthwhile to apply the proposed criterion to the image reconstruction used in each of these missions. Thus it could be determined whether these recent missions are susceptible to the same calibration issues when detecting the reported anisotropies as WMAP. Or immune to them as are the repetitive circular scans of Planck.

Extensive repetitive circular scans are being acquired as part of Planck's initial on orbit check out. As this check out is ongoing at the time of writing, anisotropy results from the Planck satellite should be available soon.

**Acknowledgements.** - Thanks to Dr. Bob W. van Dijk, Dr. Francesco Sylos Labini, Dr. Theo Nieuwenhuizen and Prof. R.M. Heethaar for their support of this work. While much of the research in this project was conducted on the author's own time, additional support was received from the Netherlands' Virtual Laboratory for e-Science (VL-e) Project, the Department of Physics and Medical Technology of the VU University Medical Center. D.W. Paty and D.K.B. LI of the MS/MRI Research Group of the University of British Columbia funded the author's initial fellowship. The DAS3 cluster at the VU University was used for the extensive computation required by this reanalysis.

Table 1. The first few calibration coefficients from the official analysis and the reanalysis for channel V11.

| Time (hours) | V113 Gain | V114 Gain | V113 Baseline | V114 Baseline |
|---|---|---|---|---|
| Calibration parameters from official analysis | | | | |
| 0.0000 | 0.449684 | -0.477935 | 19048.188 | 19122.934 |
| 0.9984 | 0.445866 | -0.488391 | 19048.115 | 19122.797 |
| 1.9968 | 0.459993 | -0.490191 | 19048.090 | 19122.827 |
| 2.9952 | 0.445503 | -0.498125 | 19048.053 | 19122.809 |
| 3.9936 | 0.440231 | -0.491096 | 19048.011 | 19122.805 |
| 4.9920 | 0.452294 | -0.479695 | 19048.043 | 19122.834 |
| Calibration parameters using the null hypothesis | | | | |
| 0.0000 | 0.442333 | -0.462057 | 19048.926 | 19123.709 |
| 0.9984 | 0.448653 | -0.471521 | 19048.860 | 19123.581 |
| 1.9968 | 0.449696 | -0.474249 | 19048.820 | 19123.619 |
| 2.9952 | 0.438652 | -0.485395 | 19048.770 | 19123.594 |
| 3.9936 | 0.431049 | -0.475227 | 19048.742 | 19123.620 |
| 4.9920 | 0.443303 | -0.464134 | 19048.758 | 19123.572 |

Table 2. Sum of the squares of the residuals for the official analysis and the proposed criterion expressed as a null hypothesis. The statistical significance of the difference between the two sum of squares is also shown in units of standard deviations.

| WMAP Channel | Summed square of residuals | | Sample Number | Statistical Significance |
|---|---|---|---|---|
| | Official | Null Hypothesis | | |
| K11 | 3148519505 | 2430242971 | 158557643 | 3722 |
| K12 | 5126155901 | 4148621415 | 158557643 | 2967 |
| Ka11 | 2761499931 | 2229053939 | 158501570 | 3007 |
| Ka12 | 4050195503 | 3277829756 | 158501570 | 2967 |
| Q11 | 8013824476 | 7447616245 | 198980008 | 1072 |
| Q12 | 2748689923 | 2603833155 | 198980008 | 785 |
| Q21 | 6758636411 | 6202579627 | 199335839 | 1266 |
| Q22 | 5790904546 | 5439689280 | 199335839 | 912 |
| V11 | 7540959177 | 7326670351 | 265030538 | 476 |
| V12 | 8096785134 | 7823561948 | 265030538 | 569 |
| V21 | 4438363283 | 4246268210 | 265030297 | 736 |
| V22 | 3876882102 | 3752532772 | 265030297 | 539 |
| W11 | 11824727975 | 11669123891 | 395243975 | 265 |
| W12 | 10575117652 | 10404159678 | 395243975 | 327 |
| W21 | 13881446455 | 13764990690 | 397776949 | 169 |
| W22 | 13796528381 | 13678810720 | 397776949 | 172 |
| W31 | 12352877867 | 12253471475 | 397766347 | 162 |
| W32 | 14554811385 | 14462875075 | 397766347 | 127 |
| W41 | 10478261545 | 10397259213 | 397328796 | 155 |
| W42 | 13161162101 | 13028552814 | 397328796 | 203 |

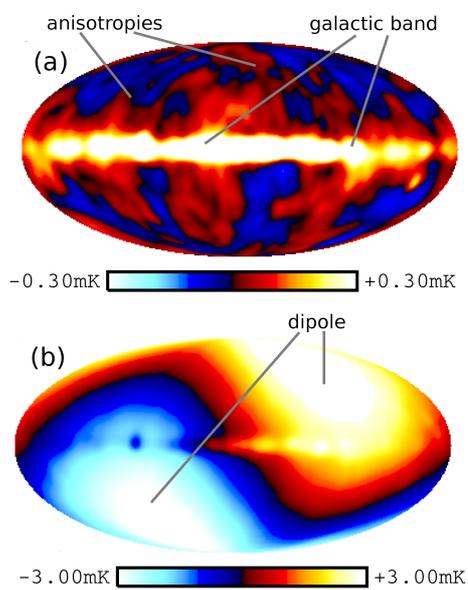

**Figure 1.** Official WMAP sky at 94GHz with (a) and without (b) the dipole subtracted.